\documentclass[preprint,12pt]{elsarticle}

\usepackage{graphicx}

\usepackage{amssymb}

\journal{Nuclear Instruments and Methods A}

\begin{document}

\begin{frontmatter}


\title{Nuclear emulsion with molybdenum filling for observation of  
$\beta\beta$ decay}


\author[ITEP]{V.D.~Ashitkov}

\author[LPI]{A.V.~Bagulya}

\author[ITEP]{A.S.~Barabash\corref{cor1}}
\ead{barabash@itep.ru}

\author[JINR]{V.Ya.~Bradnova}

\author[LPI]{M.M.~Chernyavsky}

\author[ITEP]{S.I.~Konovalov}

\author[LPI]{N.M.~Okat'eva}

\author[LPI]{G.I.~Orlova}

\author[LPI]{N.G.~Polukhina}

\author[ITEP]{E.A.~Pozharova}

\author[ITEP]{V.A.~Smirnitsky}

\author[LPI]{N.I.~Starkov}

\author[LPI]{M.S.~Vladimirov}

\author[ITEP]{V.I.~Umatov}

\cortext[cor1]{Corresponding author}

\address[ITEP]{Institute of Theoretical and Experimental Physics, B.\
Cheremushkinskaya 25, 117218 Moscow, Russian Federation}

\address[JINR]{Joint Institute for Nuclear Research, 141980 Dubna, Russia}

\address[LPI]{Lebedev Physical Institute, Leninsky prospect 53, 119991 Moscow, Russia}

\begin{abstract}

The usage of nuclear emulsion with molybdenum filling for observation of  
$\beta\beta$ decay are shown to be possible. Estimates for 1 kg of $^{100}$Mo
with zero background give the sensitivity for the $0\nu\beta\beta$ decay of
$^{100}$Mo at the level of $\sim 1.5\cdot 10^{24}$ y for 1 year of measurement.

\end{abstract}


\end{frontmatter}


\label{}

\newpage

Nuclear emulsion was first used as a simple counter of electrons to search for double 
beta decay  in 1952 \cite{FRE52}. 
In $1987-1990$ an experiment was performed searching for the $\beta\beta$ decay 
of $^{96}$Zr and $^{94}$Zr using the nuclear emulsion BR-2 and the best limits on the
$\beta\beta$ decay of $^{96}$Zr and $^{94}$Zr were obtained for these isotopes
\cite{BAR87,BAR88,BAR88a,BAR90,BAR90a}. As a result 
of those studies it became obvious that a full-scale $\beta\beta$ decay experiment 
would require an automatic scanning and much higher (by several orders of magnitude) 
scanning rates. In recent years significant progress has been made to automate 
the procedure and to increase nuclear emulsion scanning rates. 
These significant technical advance has been made possible by the OPERA 
experiment \cite{AGA09} and the creation of a fully automated 
PAVIKOM facility \cite{ALE04} at the Lebedev Physical Institute. This triggered several 
proposals to search for $\beta\beta$ decay \cite{DRA08} and dark matter 
\cite{NAK08} using nuclear 
emulsions. The main advantage of this approach in $\beta\beta$ decay studies is 
the visualization of candidate events and the possibility to measure all decay 
characteristics: the total energy, 
the energy of single electrons and the angle between the two electrons.
 
We have conducted a special investigation to elucidate the possibility of using nuclear 
emulsions in full-scale $\beta\beta$ decay experiments. A nuclear emulsion 
with very fine powder of natural molybdenum as a filler (average size of granules, 
 $\sim 2-4~\mu$m) was used. The method was refined using commercial powder to be replaced 
subsequently by a pure $^{100}$Mo source. Estimates indicate that 
molybdenum should not "spoil" the emulsion and with an optimal amount of fine powder, 
would not significantly interfere with scanning and measurements in emulsion layers. 

The molybdenum powder was introduced into nuclear emulsion during its production 
at the SLAVICH Ltd (Pereslavl-Zalessky, Russia). 
Ten plates, 9 cm by 12 cm in size, had been 
fabricated with a 75 $\mu$m thick emulsion layer  of dry substance into which 1.43 g 
of molybdenum powder ($\sim 6\%$ by weight of dry emulsion) was added. The plates were 
developed by a standard method at the JINR (Dubna, Russia). 
A KSM microscope was used to scan the plates, to measure the size of powder particles
 and to localize them in the emulsion layer. Fig. 1 shows a micrographs of 
molybdenum particles in the emulsion layer, and Fig. 2 presents their distribution 
by size. These data indicate that $\sim 80\%$ of the powder particles are less 
than 8 $\mu$m. 
 The shaded part of the histogram in Fig. 2 corresponds to the powder particles near the
bottom of the plate. Fig. 3 shows the distribution of particles as a function of
 the depth 
of the emulsion layer. The dashed line indicates the number of particles
 less than 8 $\mu$m in size in the 10 $\mu$m layer at the bottom of the plate. Thus,  
a major excess of particles at the bottom of the plate can be explained by 
the precipitation of particles greater than 8 $\mu$m during the emulsion 
coating process. The results presented in this figure suggest that there is a 
slight gradient in the distribution of molybdenum particles along the depth of 
the emulsion layer. Most likely, however, it will not interfere with the search for 
$\beta\beta$ decay in the experiment. A visual assessment shows that the amount of 
molybdenum can be increased by a factor of $\sim (1.5-2)$. This increase 
would make it possible 
to reduce the amount of emulsion in the experiment by the same factor.

\begin{figure}
\begin{center}
\includegraphics[width=8cm]{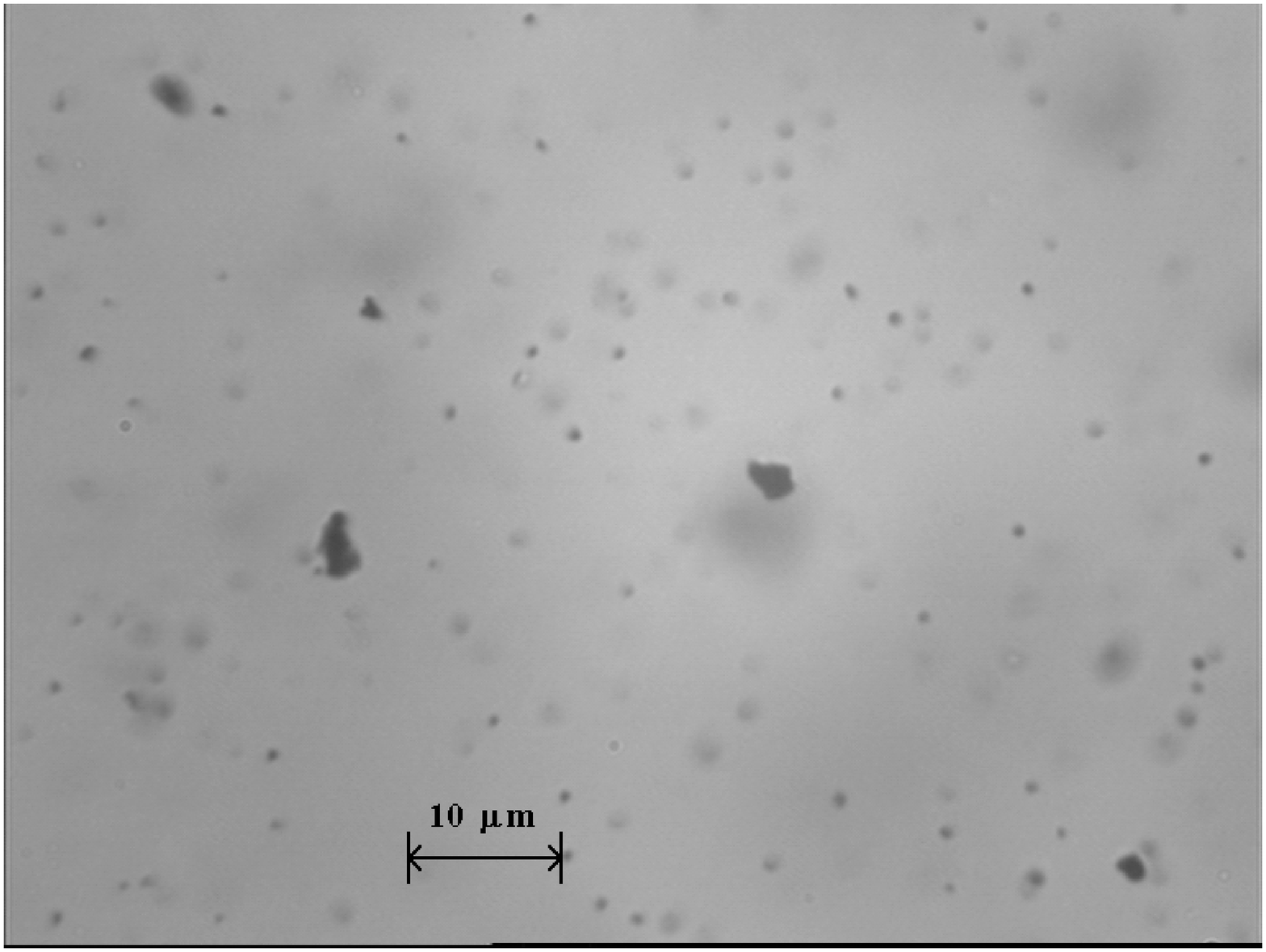}
\vspace{0.5cm}
\includegraphics[width=8cm]{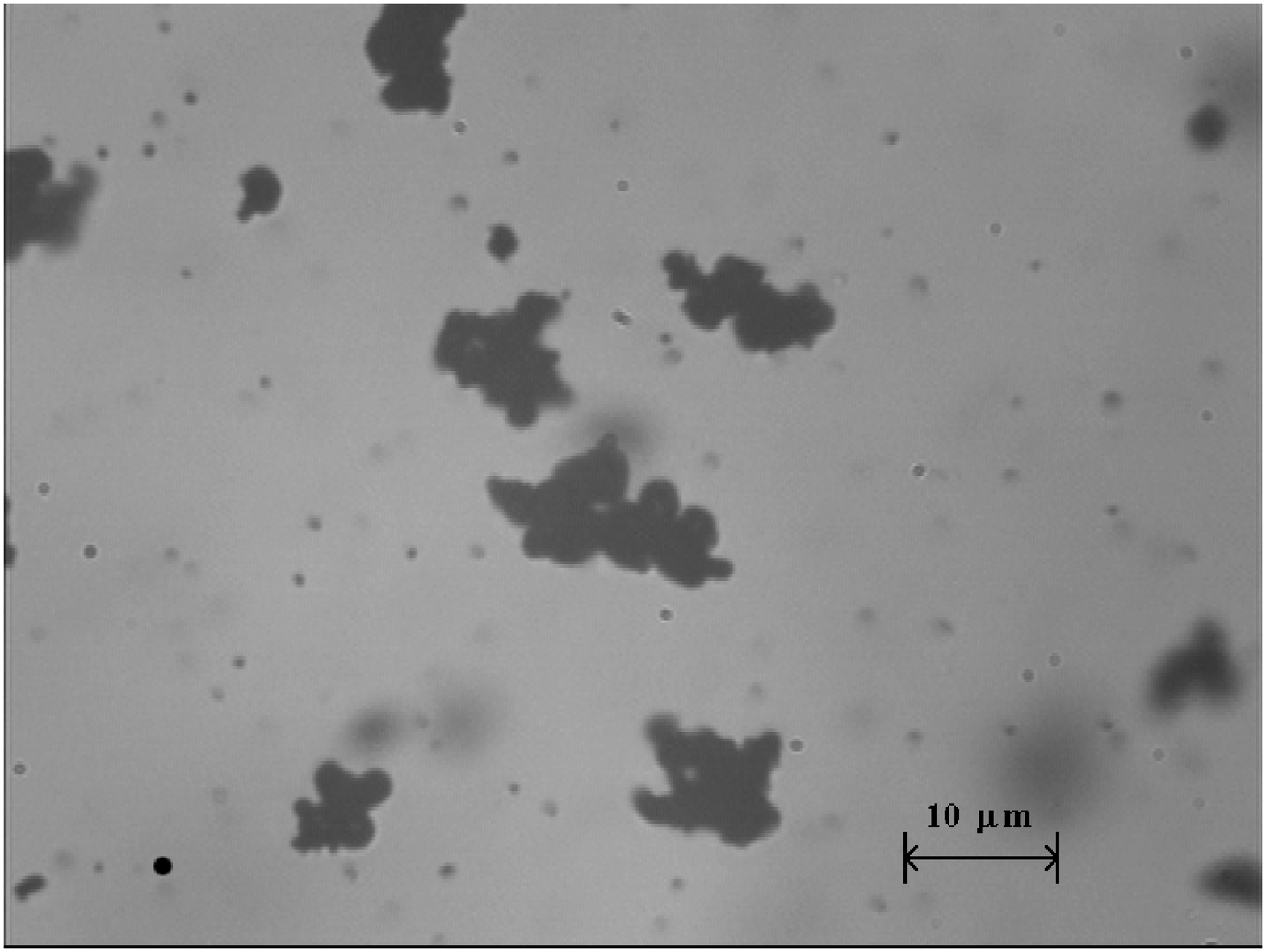}
\caption{Micrographs of the emulsion with Mo powder near the top of the plate and at the 
bottom. Definition in depth is 1 $\mu$m.}  
\label{fig_1}
\end{center}
\end{figure}

\begin{figure}
\begin{center}
\includegraphics[width=6cm]{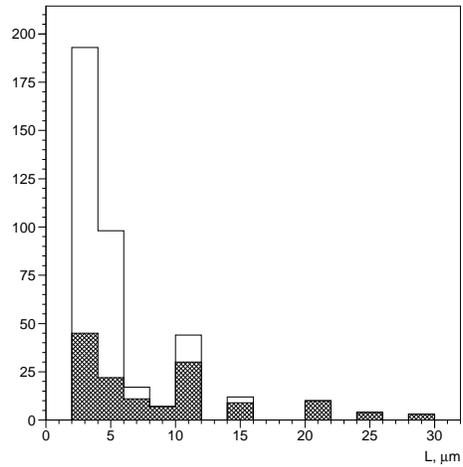}
\caption{The distribution of Mo granules versus their sizes L in the emulsion plate.
The shaded histogram
corresponds to Mo granules precipitating to the bottom layer of 10 $\mu$m during gel 
polymerization.}  
\label{fig_2}
\end{center}
\end{figure}

\begin{figure}
\begin{center}
\includegraphics[width=6cm]{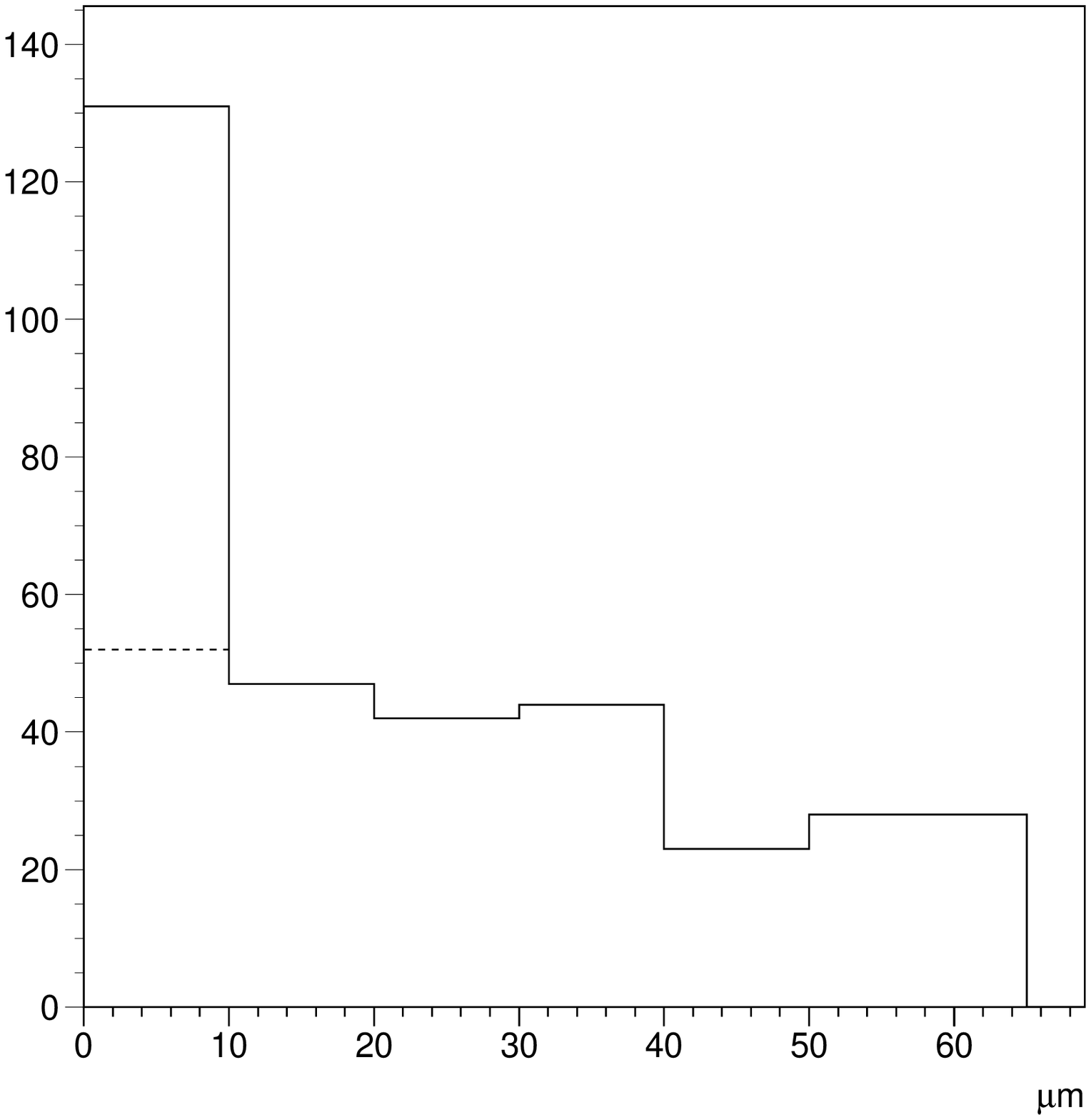}
\caption{The distribution of Mo granules from the bottom to the top of the emulsion plate 
 after development.
The dashed line corresponds to the number of Mo granules up to 8 $\mu$m 
observed in 10 $\mu$m bottom layer.}  
\label{fig_3}
\end{center}
\end{figure}

  Let us try to estimate the amount of emulsion required for a full-scale experiment.  
Assuming that the volume of the dry emulsion in our study is 8.1 cm$^3$, for 1 g of molybdenum 
the required volume of dry emulsion is 5.6 cm$^3$.  For a full-scale experiment, 
it is advisable to use $\sim 1$ kg of molybdenum in 5.6 liters 
(21.3 kg) of dry emulsion. It is known that 11.5 kg of gel is required to 
prepare 1 liter of 
dry emulsion (Ilford (Great Britain) or ET-7D (Japan)). Hence, we have: \\
\hspace{0.5cm} $\bullet$ 1 kg of molybdenum in a $\beta\beta$ decay  
experiment requires 65.5 kg of gel, which would yield 5.6 liters of dry emulsion 
or 860 plates (9x12x0.06) cm$^3$ in size; \\
\hspace{0.5cm} $\bullet$ if the content of molybdenum in emulsion is increased by a factor of 1.5-2  
then $\sim (570-430)$ emulsion layers with a thickness of $\sim 600 ~\mu$m is  
enough to assemble $10-12$ emulsion chambers; \\
\hspace{0.5cm} $\bullet$ the optimal size for granules is $\sim 2-5 ~\mu$m (if necessary, 
the molybdenum powder can be sieved before it is added to the gel); \\
\hspace{0.5cm} $\bullet$ the scanning rate at one PAVIKOM unit is $\sim 1-2$ plates 
per day, 
which should make it possible to complete scanning within one year. \\

In this case, with a zero background and one-year measurements we can achieve 
a sensitivity to 0$\nu$-decay of $^{100}$Mo at the level of $\sim 1.5\cdot 10^{24}$ years.
 This sensitivity is comparable with the result of the NEMO-3 experiment, 
T$_{1/2}(0\nu) > 1.1\cdot 10^{24}$ years \cite{TRE09} (for 7 kg of $^{100}$Mo in 
3.7 years of 
measurements). 

The success of the experiment largely depends on the accuracy of measuring the
electron energy in the nuclear emulsion. The energy of electrons is determined 
by their ranges in the emulsion chamber. The coordinates of each grain are measured, 
and the range is calculated as a sum of segments of a broken line. 
As the energy of particle with unit charge 
depends on its range as $E \sim R^{0.58}$ the representation of its trajectories 
by a broken line with the emulsion sensitivity of $\sim 30$ 
grains/$100 \mu$m yields 
an error of no more than $2-3$\%. 
Other sources of uncertainties \cite{BON,POW} come from a) the straggling,
$\sim 3$\%; b) a correction for bremsstrahlung, $\sim 1$\%; 
c)   layer-to-layer transitions (depend on angles), $\sim 3$\%.
The total error is expected to be $\sim$ 6\%. \\

The first stage of a full-scale experiment will involve an exposure with 
100 grams of $^{100}$Mo. This will enable us 
to verify the method, to study possible sources of background and to 
observe $\sim 4000 ~(2\nu\beta\beta$) decays of $^{100}$Mo for one month of exposure 
which will also be an important experimental result on its own. 
The experiment will be carried out under low-background conditions 
with an exposure time of several months. In order to accomplish this 
the nuclear emulsion should 
have an insignificant regression of a hidden image, and the sensitivity of 
emulsion layers should be checked periodically. 

Promising elements for the double beta decay search, besides molybdenum ($^{100}$Mo),
 are also $^{82}$Se, $^{150}$Nd, $^{96}$Zr, $^{130}$Te, $^{116}$Cd, and $^{48}$Ca.\\







\end{document}